\documentclass{article}
\usepackage{spconf,amsmath,graphicx}

\usepackage{graphicx}
\usepackage{amsmath}
\usepackage{amssymb}
\usepackage{color}
\usepackage{bm}

\usepackage{algorithm}
\usepackage{algpseudocode}
\algnewcommand\algorithmicinput{\textbf{Input:}}
\algnewcommand\INPUT{\item[\algorithmicinput]}
\algnewcommand\algorithmicoutput{\textbf{Output:}}
\algnewcommand\OUTPUT{\item[\algorithmicoutput]}

\usepackage{mathtools}

\newcommand{\Fig}[1]{Fig.~\textup{\ref{#1}}}
\graphicspath{{figures/}}
\usepackage{subcaption}

\newcommand{\vb}{\bm}
\newcommand{\mb}{\bm}
\newcommand{\mbe}{\bm}

\newcommand{\ksi}{\xi}

\renewcommand{\epsilon}{\varepsiolon}

\title{THEORETICAL PERFORMANCE BOUND \\
OF UPLINK CHANNEL ESTIMATION ACCURACY IN MASSIVE MIMO}
\name{Alexander Osinsky, Andrey Ivanov, Dmitry Yarotsky \thanks{The research was carried out at Skoltech and supported by the Russian Science Foundation (project no. 18-19-00673).}}
\address{Skolkovo Institute of Science and Technology, Russia \\
alexander.osinsky@skoltech.ru, an.ivanov@skoltech.ru, d.yarotsky@skoltech.ru}
\begin{document}
%\ninept
%
\maketitle
\begin{abstract}
In this paper, we present a new performance bound for uplink channel estimation (CE) accuracy in the Massive Multiple Input Multiple Output (MIMO) system. The proposed approach is based on noise power prediction after the CE unit. Our method outperforms the accuracy of a well-known Cramer-Rao lower bound (CRLB) due to considering more statistics since performance strongly depends on a number of channel taps and power ratio between them. Simulation results are presented for the non-line of sight (NLOS) 3D-UMa model of 5G QuaDRiGa 2.0 channel and compared with CRLB and state-of-the-art CE algorithms.
\end{abstract}
\section{INTRODUCTION}

The fifth generation (5G) of wireless systems will demand more users with much higher overall capacity. In recent years, massive multiple input multiple output (MIMO) has been adopted as one of the key technologies to address the capacity requirements of the enhanced Mobile Broadband (eMBB) in $5$G as described in \cite{A3}. Spatially multiplexed multi-user (MU-MIMO) systems can support several independent data streams, resulting in a significant increase of the system throughput. Some challenging issues are still unresolved in multi-antenna orthogonal frequency-division multiplexing (OFDM) systems, and one of them is channel estimation (CE) accuracy in Massive MIMO. The number of antennas in Massive MIMO starts from $64$ while in $4$G with a common MIMO this number is limited by $8$ \cite{A31}. With a growing number of antennas correlation between them gets higher, which provides extra abilities to enhance performance significantly via joint processing as shown in \cite{A20,A21,A19}, but practical performance is still too far from the performance, achieved with ideal CE. As a result, CE topic attracts many researches to compensate gap between ideal and practical CE performances \cite{A5}. Most of the existing theoretical lower bounds for CE accuracy are based on the Cramer-Rao lower bound (CRLB) \cite{A6,A7,A8}. However, this approach has several disadvantages:

1) The estimator should be unbiased. Although there is a generalization called Bayesian CRLB \cite{A6,A7,A8}, the simple version is usually used. Therefore, any use of minimum mean squared error (MMSE) can theoretically surpass such bounds;

2) The bounds do not take into account possibly different distribution of peaks and correlation between them. For example, in \cite{A6,A7,A8} peaks are considered to have arbitrary power and to be independent;

3) The bounds do not take into account the correlation between antenna elements;

Our approach solves all of these problems. The proposed CE model is based on the following assumptions:

1) The channel can be approximated by the finite number of taps;

2) The signal distribution is the same on each antenna;

3) Tap amplitudes distribution can be approximated by the multivariate Gaussian distribution.

The first two assumptions come from the physics of signal propagation. The other is also reasonable: one would usually want from the detector to be able to distinguish between the close signals and the Gaussian distribution is the most straightforward choice to provide such signals. It is well known that for Gaussian distribution of the signal and Gaussian noise MMSE estimator is the best possible one \cite{A2}. Provided all the information about the distribution, it allows constructing the best possible estimator quite easily. 

In this paper, we propose a new algorithm of residual error power estimation after CE unit for each scenario. When having the CE variance for each case, we generate noisy channel as a sum of the ideal channel and white noise with the appropriate noise power. Then this artificial CE is employed by MIMO detector and decoder units to achieve CE performance bound in the full $5G$ receiver simulator. Frame error rate (FER) is performance metric, and FER=1 for the current time transmission interval (TTI) if we have at least one false decoded bit in the information transport block ($12$ data symbols) decoding. We show FER curves for Ideal CE (channel without noise), Theoretical CE (sum of the ideal channel and the theoretical residual noise) and practical CE achieved by the state-of-the-art algorithms \cite{A20}. Theoretical performance bound could tell us, how much receiver performance gain (in dB) one can really achieve from CE algorithm in defined channel scenarios of Massive MIMO. We will be showing the decoder FER instead of the Euclidean norm of the MIMO detector error \cite{A2, A1, A10}, because that is what we are actually interesting to decrease.

\section{SIGNAL MODEL}
In practice, one distributed scatterer (ground, building and so on) is represented by infinite number of local scatterers, which can be approximated (sampled) by a finite sum of global scatterers as shown in \cite{A20}. Assume a channel model, consisting of $M$ taps. Consider channel impulse response $h\left(k,t\right)$ in the single antenna is given by a finite number of global scatterers:
\[h\left(k,t\right)=\sum^M_{m=1}{x_m\left(k\right)\delta \left[t-{\tau }_m\left(k\right)\right]},\] 
where $k$ is the antenna index; ${\tau }_m\left(k\right)$ is the delay of tap; $x_m\left(k\right)$ is the tap complex amplitude, $m$ is the tap (or beam) index, $M$ is the number of taps in the channel (or beams in multi-antenna scenario). In a band-limited case ($N_{used}<N_{DFT}$) the discrete channel impulse response can be calculated via convolution as:
\[s\left(k,\ n\right)=h\left(k,nT\right)*sinc\left(\pi \frac{N_{used}}{N_{DFT}}n\right),\]
\[s\left(k,\ n\right)=\sum^M_{m=1}{x_m\left(k\right)sinc\left(\pi \frac{N_{used}}{N_{DFT}}\left[n-\frac{{\tau }_m\left(k\right)}{T}\right]\right)},\] 
where $T$ is the sample clock period of Discrete Fourier Transform (DFT) unit, $N_{DFT}$ is the DFT size, $n \in [1...N_{DFT}]$ is the sample index; $N_{used}=N_{RB} \times RB_{size}$ is the number of utilized subcarriers in the spectrum, $N_{RB}$ is the number of allocated resource blocks, $RB_{size}=12$ is the resource block size, $N_{DFT}$ is the DFT size and  $sinc\left(x\right)=sin\left(x\right)/x$. Function $s\left(n\right)$ is called Zakai's series in the sampling theory literature \cite{A2}, where amplitudes $x_m\left(k\right)$ are defined as deterministic and bounded. If the max distance between antenna elements, divided by the speed of light ($d_{max}/c$) is much smaller than $\left(TN_{DFT}/N_{used}\right)$ value, the delay value can be assumed as independent of antenna index  $k$, i.e. ${\tau }_m\left(k\right)={\tau }_m$. 
It should be noticed that the complex amplitude value $x_m\left(k\right)$ strongly depends on antenna index $k$. Therefore, channel model is given by:
\begin{equation}\label{LS_CE}
  s\left(k,\ n\right)=\sum^M_{m=1}{x_m\left(k\right)sinc\left(\pi \frac{N_{used}}{N_{DFT}}\left[n-n_m\right]\right)},\
\end{equation}
%\[s\left(k,\ n\right)=\sum^M_{m=1}{b_m\left(k\right)sinc\left(\pi \frac{N_{used}}{N_{DFT}}\left[n-n_m\right]\right)}\]
where $n_m=round\left(\frac{{\tau }_m}{T}\right)$ is the discrete delay.

\section{BEAM DOMAIN REPRESENTATION}
Equation (\ref{LS_CE}) leads to the fact that the time domain least squared (LS) channel estimation can be described by a matrix $\mb{S} \in \mathbb{C}^{N_{DFT} \times N_{RX}}$, where $N_{RX}$ is the number of antennas. The first assumption means that this matrix can be represented as the following product:
\begin{equation}\label{eq-beam-dec}
  \mb{S} \approx \mb{B} \mb{X_0}, \quad \mb{B} \in \mathbb{C}^{N_{DFT} \times M}, \quad \mb{X_0} \in \mathbb{C}^{M \times N_{RX}},
\end{equation}
where $M$ is the maximum number of beams (taps) in the channel model, each column of $\mb{B}$ is the $sinc$ function, shifted in time to the channel tap (or beam delay) position, $\mb{X_0}$ is the matrix, containing $M$ tap amplitudes for each of $N_{RX}$ receiver antennas. Naturally, achieved beams are non-orthogonal, and to orthogonalize them we apply $QR$ decomposition to matrix $\mb{B}$ as:
\[
  \mb{B} = \mb{Q} \mb{R}, \quad \mb{Q} \in \mathbb{C}^{N_{DFT} \times M}, \quad \mb{R} \in \mathbb{C}^{M \times M}
\]
Matrix $\mb{Q}$ now describes the orthogonal subspace of the matrix $\mb{B}$. Therefore, equation $\mb{X} = \mb{R} \mb{X_0}$ describes beam amplitudes in the new subspace. Hereinafter we work only with the matrix $\mb{X}$.

Suppose an ideal channel estimator can utilize the knowledge of variance and correlation. Then it should provide average estimates for signals with the same distribution. 

\section{A NEW BOUND}
Let us remind that we work with the matrix $\mb{X} \in \mathbb{C}^{M \times N_{RX}}$, consisting of beam amplitudes, where $M$ is the number of (now orthogonal) beams and $N_{RX}$ is the number of antennas. Each row of $\mb{X}$ describes the signal for a particular beam. Moreover, the error variance $\sigma^2$ at each beam is not rescaled because beams form an orthogonal subspace. Hereinafter we consider the noise to be random Gaussian and completely uncorrelated.

For now, let us assume for simplicity that the beam amplitudes on different antennas are independent. The appropriate changes will be discussed after dealing with this simple case.  

Define noisy beam amplitudes as:
\begin{equation}\label{eq:y=x+e}
  \mb{Y} = \mb{X} + \mb{E},
\end{equation}
where matrix $\mb{E} \in \mathbb{C}^{M \times N_{RX}}$ describes the noise values in a beam subspace. We remind that the noise is random Gaussian with the average power $\sigma^2$ on each antenna:
\begin{equation}\label{eq:eij}
  \mbe{E_{ij}} \sim {\cal N} \left(0, \sigma^2 \right)
\end{equation}  

Our goal is to find the best estimate of the matrix $\mb{X}$ from $\mb{Y}$. Suppose we know the expected power of elements in $\mb{X}$ and correlation between them:
\begin{subequations}\label{eq-cor-mat}  
\begin{align}
  \mathbb{E} \left| \mbe{X_{ij}} \right|^2 & = \mbe{C_{ii}} \\
  \mathbb{E} \left( \mbe{X_{ij}}, \mbe{X_{kj}} \right) & = \mbe{C_{ik}}
\end{align}
\end{subequations}

If these statistics are not given, they can be estimated from a single sample of $\mb{X}$, using the fact that univariate distributions on each antenna are equivalent (independence is not required here):
\[
  \mb{C} = \mb{X}\mb{X}^*/{N_{RX}}.
\]
In case of independent values on different antennas
\begin{equation}\label{eq-cor-mat2}
  \mathbb{E} (\mbe{X_{ij}}, \mbe{X_{kl}}) = 0, \; j \neq l
\end{equation}
For $N_{RX} = 64$ we have $64$ samples to determine beam correlation coefficients, while the number of powerful beams in practice of $5G$ is never greater than $8$ as defined in \cite{A31}.

\emph{•}Now we know a correlation matrix $\mb{C} \in \mathbb{C}^{M \times M}$ for the beam amplitudes and the correlation matrix $\sigma^2 \mb{I_M} \in \mathbb{C}^{M \times M}$ for the noise values. Therefore, the linear observation process is given by:
\[
  \mb{Y} = \mb{A}\mb{X} + \mb{E},
\]
where $\mb{A} = \mb{I}$. Assuming the Gaussian distribution of $\mb{X}$, the best estimator is known to be linear MMSE estimator \cite{A2}. In our case it is written as follows:
\begin{equation}\label{sol-eq}
  \mb{\hat X} = \left(\mb{I_M} + \sigma^2 \mb{C}^{-1} \right)^{-1} \mb{Y}
\end{equation}
Let us remind that beam amplitudes in different antennas are still considered independent. Using (\ref{sol-eq}), the error correlation matrix can be calculated as:
\[
  \mb{C_{er}} = \mb{C} - \mb{C} \left(\mb{I_M} + \sigma^2 \mb{C}^{-1} \right)^{-1} = \sigma^2 \left( \mb{I_M} + \sigma^2 \mb{C}^{-1} \right)^{-1}
\]
Therefore, the expected total leftover noise power (expected residual squared error after CE unit) in each antenna is equal to $tr( \mb{C_{er}})$ and can be calculated as:
\begin{equation}\label{eq-noise-left}
  \frac{\mathbb{E} \left\| \mb{X} - \mb{\hat X} \right\|_F^2}{N_{RX}} = \operatorname{tr}((\sigma^{-2}\mb{I_M}+\mb{C}^{-1})^{-1}).
\end{equation}
In our simulations we do not calculate the expected beam amplitudes $\mb{\hat X}$. Instead of using equation (\ref{sol-eq}), we feed artificially noised channel response to the MIMO detector and compute FER. Namely, we use the equation (\ref{sol-eq}) with randomly generated white noise $\mb{E}$ as a CE, which is further utilized in MIMO detector by the same way as any practical CE. We note, that though equation (\ref{eq-noise-left}), like CRLB, provides us with the least possible leftover noise power, it does not guarantee the minimum possible FER after Low Density Parity Check (LDPC) decoder \cite{A14, A15, A18}.

Let us now describe how to incorporate antenna correlations into (\ref{sol-eq}). We will use two simple correlation models: ideal phase correlation and full correlation. In ideal phase correlation the estimator knows the values as:
\[
  \phi_{kj} = \arg \left(\mbe{X_{kj}} \right)
\]
Fixing arguments means that the distribution of $\mb{X}$ can be written as:
\[
  \mbe{X_{kj}} = |\ksi_{kj}| e^{i \phi_{kj}}, \quad \ksi_{kj} \in {\cal N} \left( 0, \sigma_k^2 \right), \quad \ksi_{kj} \in \mathbb{R},
\]
where $\ksi_{kj}$ are now real random variables. Their correlation matrix can be found from equations (\ref{eq-cor-mat}).
This means that the orthogonal to the signal part of the error $\mb{E}$ does not play any role: if we multiply each element of $\mb{X}$ by $e^{-i \phi_{kj}}$ we get no imaginary part of the signal and thus any detected imaginary part can be set to zero. Therefore, our problem becomes real instead of complex and $\mbe{E_{ij}}$ can then be treated as real Gaussian noise with twice smaller power. This is equivalent to keeping $\mb{X}$ the same and projecting the noise as follows:
\begin{equation}\label{eq-e-phase}
  \mbe{E_{kj}} := \operatorname{Re} \left( \mbe{E_{kj}} e^{-i \phi_{kj}} \right) \cdot e^{i \phi_{kj}},
\end{equation}
after which we use the same MMSE estimate (\ref{sol-eq}). Note, however, that now we cannot say that it indeed is the best possible one, because values of $\mbe{X_{kj}} e^{-i \phi_{kj}}$ are always positive and generally the ideal estimator should use this knowledge. On the other hand, as the noise power decreases, the probability of getting the wrong sign decreases exponentially, therefore the MMSE estimate still should be close to ideal for sufficiently small noise power.

Let us now consider full antenna correlation. It stands for the signal in each beam $\vb{x} \in \mathbb{C}^{N_{RX}}$ (row in $\mb{X}$) to have the following distribution:
\[
  \vb{x} = \eta \vb{x_0}, \quad \eta \sim {\cal N} \left(0, 1 \right), \quad \vb{x_0} \in \mathbb{C}^{N_{RX}} = const
\]
and the ideal estimator knows the value of $\vb{x_0}$.
Then the whole matrix $\mb{X}$ can be defined as:
\begin{equation}\label{eq-corr-full}
  \mbe{X_{ij}} = \eta_i \vb{x_0^j}, \quad \eta_i \sim {\cal N} \left(0, 1 \right), \quad \vb{x_0^j} \in \mathbb{C}^{N_{RX}}
\end{equation}
with the correlations of $\eta_i$ chosen to satisfy equations (\ref{eq-cor-mat}).

Let us look again at some row $\vb{x}$ of $\mb{X}$. If we project the noise on $\vb{x_0}$, the projection can be calculated as:
\begin{equation}\label{eq-e-full}
  \mb{E} := \mb{E} \vb{x_0} \vb{x_0}^* / \left| \vb{x_0} \right|^2
\end{equation}
Therefore, instead of $N_{RX}$-dimensional problem we have 1-dimensional. Nevertheless, the MMSE solution for this problem is still described by equation (\ref{sol-eq}).

Equation \eqref{sol-eq} gives us the best estimate in the orthogonal subspace. The best estimate $\mb{\hat X_0} \in \mathbb{C}^{M \times {N_{RX}}}$ of $\mb{X_0}$ in the original subspace is named {\bf Bound 1 (uncorrelated)} and can be obtained as follows:
\begin{equation} \label{eq-uncor}
  \mb{\hat X_0} = \mb{R}^{-1} \left(\mb{I_M} + \sigma^2 \mb{C}^{-1} \right)^{-1} (\mb{R}\mb{X_0} + \mb{E}) 
\end{equation}
Finally, because we are given two pilot signals, we just put $\mb{E}/\sqrt{2}$ instead of $\mb{E}$ in equation (\ref{eq-uncor}). 

In case when there is phase correlation, i.e. {\bf Bound 2 (phase correlated)} case, each element of $\mb{E}$ should be projected on the phase direction of the corresponding element of $\mb{X}$ using equation (\ref{eq-e-phase}). In the case of complete correlation, i.e. {\bf Bound 3 (fully correlated)} case, each column of $\mb{E}$ is projected on the corresponding column of $\mb{X}$ using equation \eqref{eq-e-full}. In all cases the value of $\sigma$ should be rescaled correspondingly. 
Equations (\ref{eq-e-phase}, \ref{eq-e-full} and \ref{eq-uncor}) construct an artificial CE which is used to achieve theoretical performance bounds.

%The block diagram of the theoretical CE calculation is presented in \Fig{fig10}. This theoretical CE is further utilized instead of practical Channel Estimation unit as shown in \Fig{fig2} to get the theoretical performance bound.
%
%\begin{figure}[htb]
%\centering
%\includegraphics[width=1.0\columnwidth]{diagram.jpg}
%\caption{
%Block diagram of theoretical CE
%}
%\label{fig10}
%\end{figure}

\section{SIMULATION RESULTS}
\begin{figure}[htb]
\centering
\includegraphics[width=0.9\columnwidth]{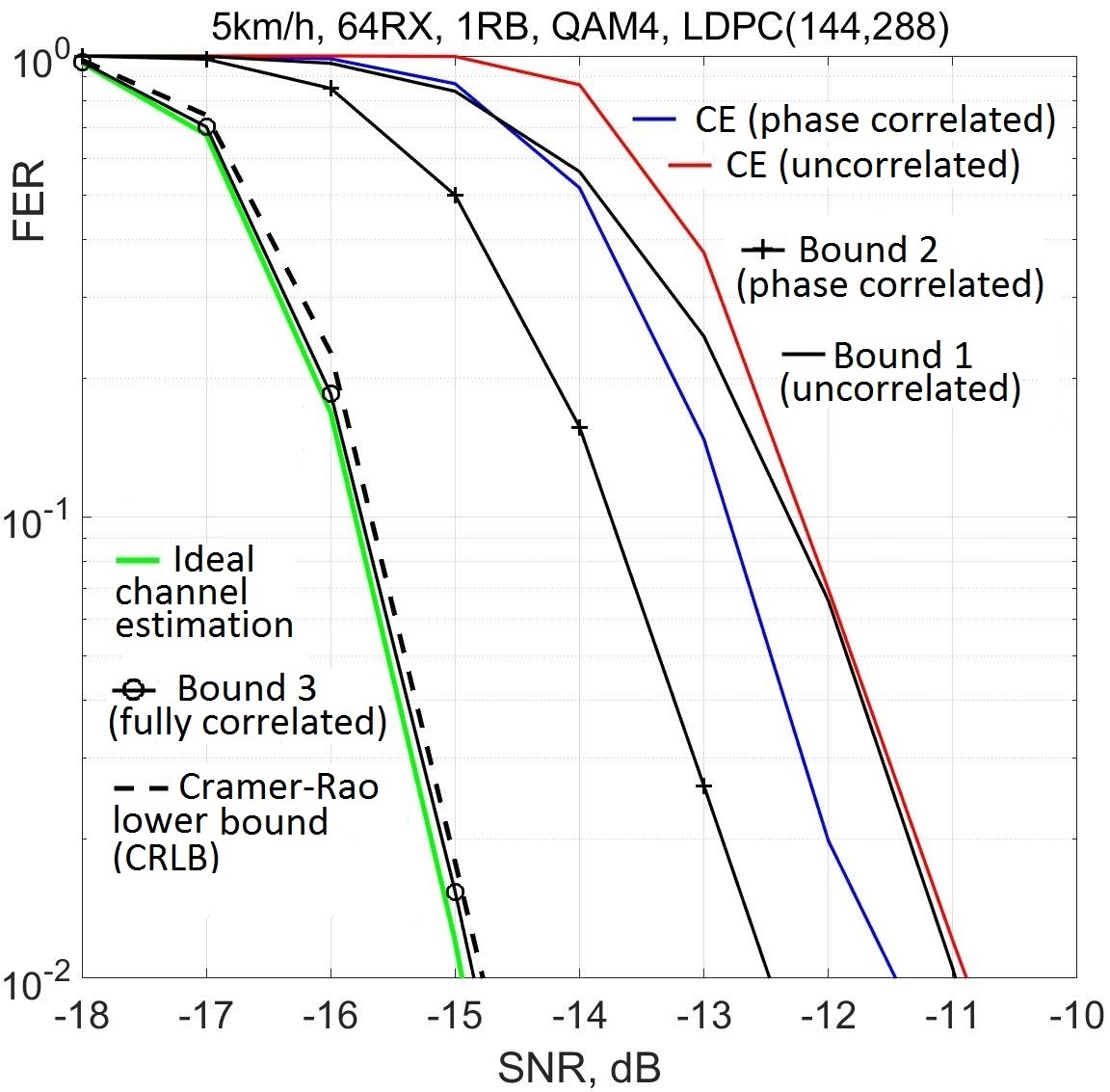}
\caption{
Performance for 1RB band
}
\label{fig7}
\end{figure}

\begin{figure}[htb]
\centering
\includegraphics[width=0.9\columnwidth]{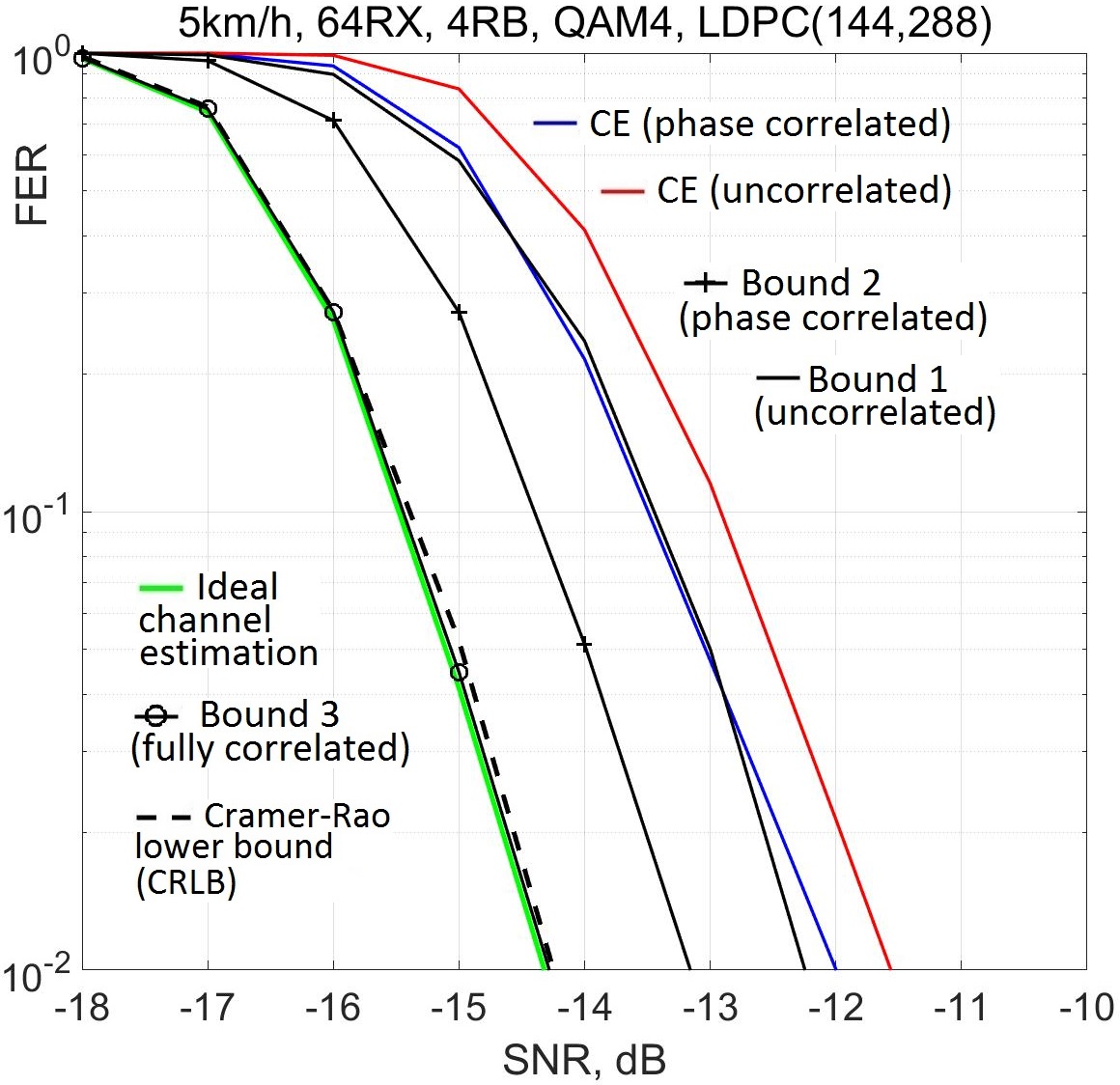}
\caption{
Performance for 4RB band
}
\label{fig8}
\end{figure}

For our simulations, we utilize non-line of sight (NLOS) 3D-UMa model of 5G QuaDRiGa 2.0 channel with $64$ antennas of the base station and single antenna user. QuaDRiGa, short for "QUAsi Deterministic RadIo channel GenerAtor" \cite{A29}, is used to generate realistic radio channel responses in simulations of mobile networks. We employed "Zhores" supercomputer \cite{A30} for parallel computing in three different models of antennas correlation: uncorrelated, phase correlated and fully correlated antennas. Results are presented in \Fig{fig7} and \Fig{fig8} for the Massive MIMO CE algorithms from \cite{A20}, theoretical bounds and ideal CE for $N_{RX} = 64$ array antennas of Massive MIMO receiver. To achieve theoretical bounds we replace standard CE unit of the 5G receiver by the artificial CE.

In case of uncorrelated antennas we suppose that antennas are not correlated at all or the CE algorithm is not allowed to use information about antennas correlation. Results are defined as {\bf Bound 1 (uncorrelated)}.

For phase correlated antennas we assume that the estimator is allowed to know the phase correlation between antennas. In practice, phase correlation can be estimated using antenna positions as shown in \cite{A151,A152}. To be able to use the same formulas as for no correlation, noise is modified using equation (\ref{eq-e-phase}). The results are shown in the figures as {\bf Bound 2 (phase correlated)}.

Finally, for fully correlated antennas case we assume all the antennas are ideally correlated, and the estimator knows the correlation exactly. To be able to use the same formulas as for no correlation, noise is modified using equation (\ref{eq-e-full}). The results are shown in the figures as {\bf Bound 3 (fully correlated)}.

State-of-the-art nonlinear CE algorithms are defined as {\bf CE (uncorrelated)} and {\bf CE (phase correlated)}. Full correlated antennas case is not analyzed since magnitude correlation between antennas is quite low in realistic NLOS channel, and its consideration does not bring extra performance gain. It can be found that practical {\bf CE (uncorrelated)} is too close to the {\bf Bound 1 (uncorrelated)} in the $1$RB scenario, i.e. theoretical limit is almost achieved. Let us remind, that {\bf CE (uncorrelated)} does not utilize any knowledge of the antennas correlation as well as {\bf Bound 1 (uncorrelated)}. Algorithm {\bf CE (correlated)} is still far from the {\bf Bound 2 (phase correlated)} because of limited beam angles estimation accuracy. 

Finally, we plot FER performance for theoretical CRLB \cite{A5, A6, A7, A8}, where the residual CE noise is generated as random Gaussian in the signal subspace. As we can see, general CRLB is poorly suited for realistic CE.

\section{ACKNOWLEDGMENT}
The authors acknowledge the use of Zhores for obtaining the results presented in this paper.

\section{CONCLUSION}
Our approach provides quite accurate lower bounds. Nevertheless, the performance gap between theoretical bounds and the practical algorithm can still be quite substantial, but less than the gap between practical algorithm and ideal CE performances. It can be explained by the fact that algorithms are not provided with the full knowledge of the distribution and have to obtain it from the noisy pilots. In particular, tap delays and directions of arrival can't be obtained exactly in the presence of noise. However, estimating an error coming from the incomplete information of channel taps is quite challenging.

\vfill\pagebreak


\begin{thebibliography}{99}

\bibitem{A3}
5G PPP Architecture Working Group,
\newblock View on 5G Architecture,
\newblock \emph {Version 3.0}, June 2019.

\bibitem{A31}
https://www.3gpp.org/release-15

\bibitem{A20}
H. Xie, F. Gao and S. Jin, 
\newblock An Overview of Low-Rank Channel Estimation for Massive MIMO Systems, 
\newblock \emph {IEEE Access}, vol. 4, pp. 7313-7321, 2016.

\bibitem{A21}
H. Al-Salihi, M. R. Nakhai and T. A. Le, 
\newblock DFT-based Channel Estimation Techniques for Massive MIMO Systems,
\newblock \emph {2018 25th International Conference on Telecommunications (ICT)}, St. Malo, 2018, pp. 383-387.

\bibitem{A19}
M. Jiang, G. Yue, N. Prasad and S. Rangarajan, 
\newblock Enhanced DFT-Based Channel Estimation for LTE Uplink,
\newblock \emph {2012 IEEE 75th Vehicular Technology Conference (VTC Spring)}, Yokohama, 2012, pp. 1-5.

\bibitem{A5}
G. T. Zhou, M. Viberg and T. McKelvey, 
\newblock A first-order statistical method for channel estimation
\newblock \emph {IEEE Signal Processing Letters}, vol. 10, no. 3, pp. 57-60, March 2003.

\bibitem{A6}
L. Berriche, K. Abed-Meraim and J. C. Belfiore, 
\newblock {Cramer-Rao bounds for MIMO channel estimation},
\newblock \emph {2004 IEEE International Conference on Acoustics, Speech, and Signal Processing}, Montreal, Que., 2004.

\bibitem{A7}
L. Berriche and K. Abed-Meraim
\newblock Stochastic Cramer-Rao bounds for semiblind MIMO channel estimation,
\newblock \emph {Proceedings of the Fourth IEEE International Symposium on Signal Processing and Information Technology}, Rome, 2004, pp. 119-122.

\bibitem{A8}
L. Berriche, K. Abed-Meraim and J. Belfiore,
\newblock Investigation of the channel estimation error on MIMO system performance,
\newblock \emph {2005 13th European Signal Processing Conference}, Antalya, 2005, pp. 1-4.

\bibitem{A2}
S.~Verdu,
\newblock Multiuser detection,
\newblock \emph {Cambridge university press}, 1998.

\bibitem{A1}
A. Ivanov, D. Yarotsky, M. Stoliarenko and A. Frolov,
\newblock Smart Sorting in Massive MIMO Detection,
\newblock \emph {14th International Conference on Wireless and Mobile Computing, Networking and Communications (WiMob)}, Limassol, 2018, pp. 1-6.

\bibitem{A10}
A. Ivanov, A. Savinov and D. Yarotsky,
\newblock Iterative Nonlinear Detection and Decoding in Multi-User Massive MIMO,
\newblock \emph {15th International Wireless Communications and Mobile Computing Conference (IWCMC)}, Tangier, Morocco, 2019, pp. 573-578.

\bibitem{A14}
R.~G.~Gallager, 
\newblock \emph{Low-Density Parity-Check Codes}.
\newblock Cambridge: MIT Press, 1963.

\bibitem{A15}
R.~Tanner,
\newblock A recursive approach to low complexity codes.
\newblock \emph{IEEE Trans. Inf. Theory}, 
vol. 27, no. 5, pp. 533--547, Sep. 1981.

\bibitem{A18}
A. Frolov,
\newblock An Upper Bound on the Minimum Distance of LDPC Codes over GF(q),
\newblock \emph {Proc. IEEE Int. Symp. Inf. Theory}, pp. 2885-2888, 2015.

\bibitem{A151}
H. Xie, F. Gao, S. Zhang and S. Jin, 
\newblock Spatial-Temporal BEM and Channel Estimation Strategy for Massive MIMO Time-Varying Systems,
\newblock \emph{2016 IEEE Global Communications Conference (GLOBECOM)}, Washington, DC, 2016, pp. 1-6.

\bibitem{A152}
J. Zhao, F. Gao, W. Jia, J. Zhao and W. Zhang, 
\newblock Channel tracking for massive MIMO systems with spatial-temporal basis expansion model,
\newblock \emph {2017 IEEE International Conference on Communications (ICC)}, Paris, 2017, pp. 1-5.

\bibitem{A29}
http://quadriga-channel-model.de/

\bibitem{A30}
I. Zacharov, R. Arslanov, M. Gunin, D. Stefonishin, A. Bykov, S. Pavlov, O. Panarin, A. Maliutin, S. Rykovanov, M. Fedorov, 
\newblock "Zhores" - Petaflops supercomputer for data-driven modeling, machine learning and artificial intelligence installed in Skolkovo Institute of Science and Technology,   
\newblock \emph {Open Engineering}. 9. 512-520. 10.1515/eng-2019-0059. 

\end{thebibliography}
\end{document}